\documentclass[
reprint,
aps,
prl,
amsmath,amssymb,
longbibliography,
floatfix
]{revtex4-2}

\usepackage{graphicx}
\usepackage{bm}
\usepackage{microtype}
\usepackage{xcolor}
\usepackage{bbm}
\definecolor{paperblue}{RGB}{0,72,153}
\usepackage{hyperref}
\hypersetup{
	colorlinks=true,
	breaklinks=true,
	linkcolor=paperblue,
	citecolor=paperblue,
	urlcolor=paperblue
}
\usepackage{tikz}
\usetikzlibrary{arrows.meta,calc,positioning}
\newcommand{\dd}{\mathrm{d}}

\newcommand{\ii}{\mathbbm{i}}

\newcommand{\Li}{\operatorname{Li}}
\newcommand{\Sec}[1]{\emph{\color{paperblue}#1.---}}

\begin{document}
	
	\title{Nonisospectral Integrability and Exact Hydrodynamic Current Fluctuations in the Two-Dimensional SSEP}
	\author{Tingfei Li}
	\email{tfli@hbu.edu.cn}
	\affiliation{College of Physics Science and Technology, Hebei University, Baoding, 071002, China}
	\affiliation{Hebei Key Laboratory of High-precision Computation and Application of Quantum Field Theory, Baoding, 071002, China}
	\affiliation{Hebei Research Center of the Basic Discipline for Computational Physics, Baoding, 071002, China}
	\date{\today}
	
	\begin{abstract}
		We study annealed current fluctuations in the two-dimensional symmetric simple exclusion process (SSEP) across a circular passive counting boundary in the joint hydrodynamic limit $T\to\infty$ and $R=r_0\sqrt{T}$ with fixed $r_0$. The initial average density is $\rho_1$ inside the disk and $\rho_2$ outside, and the observable is the net decrease of the particle number in the disk over the observation time $T$.  Within macroscopic fluctuation theory, we
		use the rotational symmetry of the problem to reduce the two-dimensional
		calculation to a radial one. A suitable change of variables then maps the
		radial problem to a nonisospectral problem on the half-line. Using the
		associated scattering construction and scalar factorization, we obtain a
		closed expression for the scaled cumulant generating function, and hence for the annealed large-deviation statistics of
		the current. Our results provide an exact benchmark for large-deviation
		phenomena in higher-dimensional interacting diffusive systems.
	\end{abstract}
	
	\maketitle
	
	\begin{figure}[t]
		\centering
		\includegraphics[width=0.70\columnwidth]{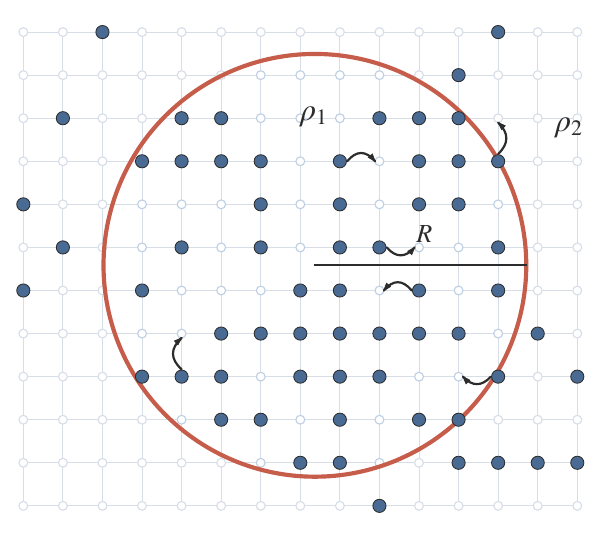}
		\caption{Radial counting geometry for the two-dimensional SSEP. A disk $B_R$ of radius $R$ separates Bernoulli densities $\rho_1$ and $\rho_2$. Its boundary is passive: particles may cross it repeatedly in either direction, while $Q_T=N_B(0)-N_B(T)$ measures the net outward transfer over time $T$.}
		\label{fig:model}
	\end{figure}
	\Sec{Large deviations and MFT}
	Large fluctuations of conserved currents reveal aspects of nonequilibrium
	transport that are not seen in average behavior or Gaussian hydrodynamic
	fluctuations \cite{Spitzer1970,Liggett1985,Spohn1991,KipnisLandim1999,
		DemboZeitouni1998,FreidlinWentzell2012,Touchette2009,
		GallavottiCohen1995,LebowitzSpohn1999}. For a current $Q_T$ with diffusive
	scaling, the scaled cumulant generating function (SCGF) and rate function are
	defined by
	\begin{align}
		\mu_d(\lambda)&=\lim_{T\to\infty}T^{-d/2}\log\left\langle e^{\lambda Q_T}\right\rangle,
		\label{eq:intro-scgf}\\
		I_d(q)&=\sup_{\lambda}\{\lambda q-\mu_d(\lambda)\}.
		\label{eq:intro-rate}
	\end{align}
	The probability of observing a macroscopic current fluctuation
	$Q_T\simeq T^{d/2}q$ obeys the large-deviation form:  $\mathbb P(Q_T\simeq T^{d/2}q)\asymp e^{-T^{d/2}I_d(q)}$. Exact
	SCGFs are rare. They usually require either an exact microscopic solution
	followed by a scaling limit or a solvable coarse-grained theory.
	
    For interacting diffusive systems, macroscopic fluctuation theory (MFT)
    describes a rare current through an optimal density--current history. It links
    hydrodynamic limits, fluctuation relations, and path-space variational
    principles \cite{Bertini2002,Bertini2003,Bertini2005PRL,Bertini2006JSP,
    	BertiniHeat2005,BertiniRMP2015,DerridaReview2007,Derrida2025Lectures,
    	BodineauDerrida2025Perturbative,BertiniGabrielliJonaLasinio2025Schrodinger,
    	SahaJangidEtAl2026BottomUp,OnsagerMachlup1953I,OnsagerMachlup1953II,
    	MartinSiggiaRose1973,Kamenev2011}. The form of $\mu_d$ can also reveal
    changes in the optimal history and dynamical phase transitions.

	\Sec{Exact results for exclusion processes}
    Exactly solvable stochastic models have played a central role in the study of
	nonequilibrium fluctuations. These models include independent particles,
	exclusion and (a)symmetric exclusion processes, zero-range processes, and
	energy-transport models \cite{Spitzer1970,Liggett1985,Spohn1991,
		KipnisLandim1999,DerridaLebowitzSpeer2002JSP,
		DerridaLebowitzSpeer2002PRL,DerridaLebowitzSpeer2003,
		BodineauDerrida2004}. Among these models, the symmetric simple exclusion
	process (SSEP) is one of the simplest and most important interacting diffusive
systems, with a density-dependent mobility. Exact results for boundary-driven
and periodic SSEP found nonlocal nonequilibrium free energies. They also motivated the
	additivity principle and its dynamical extensions
	\cite{DerridaLebowitzSpeer2002JSP,DerridaLebowitzSpeer2002PRL,
		DerridaLebowitzSpeer2003,EnaudDerrida2004,DerridaDoucotRoche2004,
		BodineauDerrida2004,BodineauDerrida2005,BodineauDerrida2007,
		LecomteEtAl2007,AppertRollandEtAl2008,HurtadoGarrido2009,
		HurtadoGarrido2011,TailleurKurchanLecomte2008,
		GorissenVanderzande2012,AkkermansEtAl2013}.

	For the one-dimensional SSEP, current fluctuations have been studied with
	Bethe ansatz, Fredholm determinants, probability theory, and MFT \cite{DerridaGerschenfeld2009JSP136,DerridaGerschenfeld2009JSP137,
		KrapivskyMeerson2012,VilenkinMeersonSasorov2014,MeersonSasorov2014,
		ProlhacMallick2008,TracyWidom2008,TracyWidom2009,
		MallickMoriyaSasamoto2022,MallickMoriyaSasamoto2024,
		DandekarKrapivskyMallick2024Dyson,BerliozBenichouGrabsch2025Driven,
		BerliozBenichouGrabsch2026Fick,SuzukiSasamoto2026,JangidEtAl2026}.
	Exact results are also available for tagged-particle and single-file
	observables \cite{Arratia1983,DeMasiFerrari2002,VandenbergRodes2010,
		SethuramanVaradhan2013,KrapivskyMallickSadhu2014,
		KrapivskyMallickSadhu2015,KrapivskyMallickSadhu2015Dyn,
		SadhuDerrida2015,ImamuraMallickSasamoto2021,GrabschEtAl2022,
		XueZhao2024,GrabschVenturelliBenichou2025PRL,
		GrabschVenturelliBenichou2026}. Recent work has additionally addressed
	dynamical phase transitions, anomalous current fluctuations, and quantum
	exclusion processes \cite{Hurtado2025Lectures,SchorleppShpielberg2025,
		YoshimuraKrajnik2025,YoshimuraKrajnik2026,AlbertEtAl2026QSSEP}.

	\Sec{Curved geometry and present work}
	Despite the many exact results available in one dimension, much less is known
	when the counting region has a curved boundary. In higher dimensions, current
	fluctuations can depend on the shape and dimension of the counting surface
	\cite{AkkermansEtAl2013,BerliozEtAl2024,GabrielliHarris2025}. Exact
	time-dependent full counting statistics are still scarce. Radially symmetric
	MFT has been applied to absorption, survival, and target problems
	\cite{MeersonVilenkinKrapivsky2014,Meerson2015Absorption,
		AgranovMeerson2017,Bressloff2025Targets}.  Semi-infinite systems,
	localized defects, and boundary-driven nonequilibrium states provide further
comparison cases \cite{GrabschEtAl2024,SahaEtAl2026,SahaSadhu2025,
CarinciEtAl2025Harmonic}. 
 
    In these problems, the surface is typically absorbing or otherwise dynamically
    active. In this Letter, we study the annealed large deviations of particle loss
    from a finite macroscopic disk in the two-dimensional SSEP. We consider a
    curved passive counting surface. It is the higher-dimensional analogue of a
    passive counting point in the one-dimensional SSEP. The surface does not alter
    the dynamics. Particles may cross it repeatedly in either direction, and the
    observable records only their net transfer. The finite curved geometry leads to
    a nonisospectral integrable structure.

	\Sec{Model and Exact Result}
	In the SSEP, particles hop randomly to neighboring sites with equal probability in both directions.  The exclusion constraint allows at most one particle at each site.  We consider the process on $\mathbb{Z}^d$ with diffusivity set to one \cite{Spohn1991,KipnisLandim1999}. The system is initially prepared in an
	annealed Bernoulli state with a spherical density profile
	\begin{equation}
		\bar\rho(\bm s)=
		\begin{cases}
			\rho_1,&|\bm s|<R,\\
			\rho_2,&|\bm s|>R .
		\end{cases}
		\label{eq:microstep}
	\end{equation}
	The initial occupation variables are sampled independently according to this
	local density. We study the net particle loss $Q_T:=N_B(0)-N_B(T)$, where
	$N_B(\tau)$ is the particle number inside the ball at physical time $\tau$, as shown in
	Fig.~\ref{fig:model}.
	
	We take the joint diffusive hydrodynamic limit $T\to\infty$, $R\to\infty$,
	with $\bm r=\bm s/\sqrt{T}$, $t=\tau/T$, and the macroscopic radius
	$r_0=R/\sqrt{T}$ held fixed.
    The rescaled current is
	$q=Q_T/T^{d/2}$, and its SCGF and rate function are defined in
	Eq.~\eqref{eq:intro-scgf} and Eq.~\eqref{eq:intro-rate}, respectively.
	
\begin{figure}[t]
		\centering
		\includegraphics[width=1.0\columnwidth]{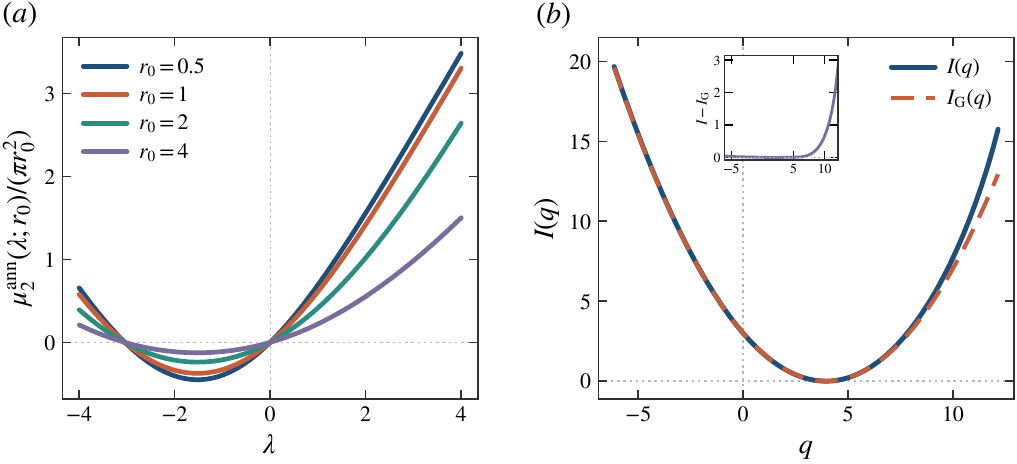}
		\caption{
			Full counting statistics for $\rho_1=0.7$ and $\rho_2=0.1$.
			(a) Area-normalized SCGF
			$\mu_2^{\rm ann}(\lambda)/(\pi r_0^2)$
			for $r_0=0.5,1,2,4$.
			(b) Exact rate function $I(q)$ at $r_0=2$
			compared with the Gaussian approximation
			$I_G(q)=(q-\bar q)^2/[2\mu_2''(0)]$,
			with $\bar q=\mu_2'(0)$.  The inset shows $I-I_G$.
		}
		\label{fig:scgf-rate}
	\end{figure}

	We now focus on the two-dimensional case.  At leading order in the
	hydrodynamic large-deviation limit, the MFT SCGF depends on the initial
	densities and the counting field only through the variable
	$\omega(\lambda)=\rho_1(1-\rho_2)(e^\lambda-1)
	+\rho_2(1-\rho_1)(e^{-\lambda}-1)$
	\cite{DerridaGerschenfeld2009JSP136,Cantini2026}. Therefore, we write
	$\mu^{\rm ann}_2(\lambda)=\mathsf F_2(\omega(\lambda))$. Our main result is
	the following MFT-exact expression:
	\begin{equation}
		\mathsf F_2(\omega)
		=
		2r_0^2
		\int_0^\pi
		\cos^2\frac{\theta}{2}
		\log\left[
		1+\omega e^{-r_0^2\sin^2(\theta/2)}
		\right]
		\,\mathrm d\theta .
		\label{eq:mainresult}
	\end{equation}
	Figure~\ref{fig:scgf-rate}(a) shows the SCGF for representative values of
	$r_0$.
	

    \Sec{Radial Optimality}
    For the annealed problem, the initial density is optimized using the
    Bernoulli relative entropy. 
    In density-current variables, the dynamical MFT cost is
    \cite{Bertini2005PRL,Bertini2006JSP,BertiniRMP2015}
    \begin{equation}
    	\int_0^1\!\dd t\int\!\dd^dr\,
    	\frac{|\bm j+\nabla\rho|^2}{4\rho(1-\rho)},
    	\qquad
    	\partial_t\rho+\nabla\!\cdot\bm j=0.
    	\label{eq:primalMFT}
    \end{equation}
    For $0<\rho<1$, the Hessian of
    $|\bm w|^2/[4\rho(1-\rho)]$ with respect to
    $(\rho,\bm w)$ is positive semidefinite. Hence the local dynamical
    integrand is jointly convex in $\rho$ and $\bm j+\nabla\rho$; the initial
    Bernoulli relative entropy is convex as well.
    Rotationally averaging any admissible history over $SO(d)$
    preserves the continuity equation, the radial reference profile,
    and $Q_T$, while Jensen's inequality~\cite{Jensen1906} guarantees that neither the
    dynamical cost nor the initial entropy increases. Hence the unrestricted MFT
    infimum equals its rotationally invariant restriction. For
    nondegenerate annealed densities the minimizer is in fact unique,
    so rotational invariance of the variational problem forces the
    physical saddle itself to be radial. A detailed proof, including
    the equality case and endpoint terms, is given in \cite{SM}.
    
    We thus consider the two-dimensional radial saddle with
    $\rho=\rho(r,t)$ and $h=h(r,t)$. The MFT equations reduce to
    \begin{align}
    	\partial_t\rho
    	&=
    	\frac{1}{r}\partial_r(r\rho_r)
    	-\frac{1}{r}\partial_r
    	\left[2r\rho(1-\rho)h_r\right],
    	\nonumber\\
    	\partial_t h
    	&=
    	-\frac{1}{r}\partial_r(rh_r)
    	-(1-2\rho)h_r^2 .
    	\label{eq:mft-h-radial}
    \end{align}
    Here and below we use the notation $A_r=\partial_r A$. The far-field
    conditions are $\rho(r,t)\to\rho_2$ and $h(r,t)\to0$ as $r\to\infty$.
    We write $A_{\rm i}(r):=A(r,0)$ and $A_{\rm f}(r):=A(r,1)$ for the
    initial and final endpoint values of any field.  The annealed temporal
    boundary conditions are
    \begin{align}
    	h_{\rm f}(r)&=-\lambda B(r),\nonumber\\
    	h_{\rm i}(r)&=
    	\log\frac{\rho_{\rm i}(r)}{1-\rho_{\rm i}(r)}
    	-\log\frac{\bar\rho(r)}{1-\bar\rho(r)}
    	-\lambda B(r),
    	\label{eq:annp0}
    \end{align}
    where $B(r)=\mathbf 1_{r<r_0}$.  The regularity conditions are
    $\rho_r(0,t)=h_r(0,t)=0$.
	
	\Sec{Two-Dimensional Radial Reduction}
	For $d=2$, we introduce the matrix field
	\begin{equation}
		\Sigma=
		\begin{pmatrix}
			1-2\rho&2(1-\rho)e^h\\
			2\rho e^{-h}&2\rho-1
		\end{pmatrix},
		\qquad
		\Sigma^2=I .
		\label{eq:Sigma}
	\end{equation}
	Using the area coordinate $s=r^2/2$ (so that $s_0=r^2_0/2$), the radial MFT equations reduce to the
	Landau--Lifshitz equation \cite{Lakshmanan1977,Takhtajan1977,LakshmananPorsezian1990,
		PorsezianLakshmanan1991}
	\begin{equation}
		\Sigma_t=[(s\Sigma_s)_s,\Sigma].
		\label{eq:radialLL2d}
	\end{equation}
	
	Only in two dimensions does this change of variables simplify both the radial
	measure and the nonlinear coefficient.  This makes the integrable structure
	below possible.

    	\Sec{Nonisospectral Lax Representation}
    Exact inverse-scattering formulations of time-dependent MFT are known for the
    one-dimensional SSEP and related diffusive or weak-noise systems
    \cite{MallickMoriyaSasamoto2022,MallickMoriyaSasamoto2024,
    	JangidEtAl2026,BerliozBenichouGrabsch2026Fick,
    	KipnisMarchioroPresutti1982,BettelheimSmithMeerson2022PRL,
    	BettelheimSmithMeerson2022JSTAT,BettelheimMeerson2024,
    	KrajenbrinkLeDoussal2021,KrajenbrinkLeDoussal2022,
    	KrajenbrinkLeDoussal2023,JanasKamenevMeerson2016,Tsai2023,
    	SchorleppSasorovMeerson2023,Bettelheim2024Whitham,Cantini2026}.
    Here the radial two-dimensional equation requires a nonisospectral
    representation.  Following the canonical spectral parametrization~\cite{SM},
    we introduce a time-independent complex label $z$ and
    \begin{equation}
    	k(t;z)=\frac{1-z}{4\ii\,[t+(1-t)z]},
    	\qquad \partial_tk=-4\ii k^2.
    	\label{eq:k-main}
    \end{equation}
    %
    Here $\ii=\sqrt{-1}$. In particular, $k(t;1)=0$ for the whole interval $0\le t\le1$, so $z=1$ is a regular base point for the half-line construction. 
     In the standard
    zero-curvature framework of inverse-scattering theory~\cite{AKNS1974,AblowitzClarkson1991,LakshmananPorsezian1990},
    the present nonisospectral Lax pair is
    \begin{align}
    	\Psi_s&=\ii k\Sigma\Psi, \nonumber \\
    	\Psi_t&=(4sk^2\Sigma-2\ii sk\Sigma\Sigma_s)\Psi.
    	\label{eq:laxt-main}
    \end{align}
    Its zero-curvature condition is precisely Eq.~\eqref{eq:radialLL2d}.  The
    evolution of $k$ compensates the explicit factor of $s$ in the radial
    equation and distinguishes the construction from the usual isospectral
    one-dimensional scattering problem.

    We introduce the transformation matrix
    \begin{align}
    	G&=\begin{pmatrix}
    		(1-\rho)e^\phi&e^{h-\phi}\\
    		\rho e^{\phi-h}&-e^{-\phi}
    	\end{pmatrix},\nonumber\\
    	\phi(s,t)&=-\int_s^\infty\rho(y,t)h_y(y,t)\dd y,
    \label{eq:frame}
    \end{align}
    which satisfies $G^{-1}\Sigma G=\sigma_3$, with
    $\sigma_3=\operatorname{diag}(1,-1)$. The columns of $G$ diagonalize the
    local matrix $\Sigma$, while the nonlocal phase $\phi$ removes the diagonal
    part of the remaining spatial connection. The resulting potential is purely
    off-diagonal and can be treated by standard half-line scattering. This
    generalizes the Cole--Hopf/inverse-scattering construction for the
    time-dependent SSEP MFT \cite{MallickMoriyaSasamoto2022} to the radial
    two-dimensional nonisospectral setting.

    Writing $\Psi(s,t)=G(s,t)e^{\ii k(t)s\sigma_3}Y(s,t)$ yields
    \begin{align}
    	Y_s&=\mathcal U(s,t )Y,\nonumber\\
    	\mathcal U(s,t )&=-
    	\begin{pmatrix}
    		0&v(s,t)e^{-2\ii ks}\\
    		u(s,t)e^{2\ii ks}&0
    	\end{pmatrix}.
    	\label{eq:A-main}
    \end{align}
    where
    $
    v(s,t)=e^{h-2\phi}h_s,
    u(s,t)=-e^{2\phi-h}[\rho_s-\rho(1-\rho)h_s]$.
    We denote the half-line transfer matrix by
    $
    \mathcal T(t )=\mathcal P\exp\!\left[\int_0^\infty
    \mathcal U(s,t )\dd s\right].
    $
    At the temporal endpoints, the annealed boundary conditions imply that
    $u_{\rm i}$ and $v_{\rm f}$ vanish away from the counting interface. Their
    respective jumps at $s=s_0$ therefore appear as localized shears \cite{SM}:
    \begin{equation}
    	u_{\rm i}(s)=\mathsf{u} \delta(s-s_0),\qquad
    	v_{\rm f}(s)=\mathsf{v}\delta(s-s_0),
    	\label{eq:delta}
    \end{equation}
    where $\mathsf{u}$ and $\mathsf{v}$ are the interface-jump strengths.
	
	\Sec{Half-Line Scattering and Scalar Factorization}
	Define the regular endpoint transforms
	\begin{align}
		\widehat v_{\gtrless}\equiv \int_{s\gtrless s_0}v_{\rm i}(s)e^{-2\ii k_{\rm i}s}\dd s,
		\widehat u_{\gtrless}\equiv \int_{s\gtrless s_0}u_{\rm f}(s)e^{2\ii k_{\rm f}s}\dd s,
		\label{eq:hatuv-main}
	\end{align}
	and the dressed interface amplitudes
	$\mathsf{u}^{\rm dr}(z)=\mathsf{u} e^{2\ii k_{\rm i}s_0}$ and
	$\mathsf{v}^{\rm dr}(z)=\mathsf{v} e^{-2\ii k_{\rm f}s_0}$.  With 
	\begin{equation}
		U(a)=\begin{pmatrix}1&-a\\0&1\end{pmatrix},\qquad
		L(a)=\begin{pmatrix}1&0\\-a&1\end{pmatrix},
		\label{eq:UL-main}
	\end{equation}
	the initial and final endpoint matrices are
	\begin{equation}
		\mathcal T_{\rm i}=U(\widehat v_>)L(\mathsf{u}^{\rm dr})U(\widehat v_<),~~
		\mathcal T_{\rm f}=L(\widehat u_>)U(\mathsf{v}^{\rm dr})L(\widehat u_<).
		\label{eq:T01-main}
	\end{equation}
	To keep track of both the time endpoint and the spatial side of the counting
	interface, define $\mathsf V_{\gtrless}:=1+\mathsf{u}^{\rm dr}\widehat v_{\gtrless}$ and
	$\mathsf U_{\gtrless}:=1+\mathsf{v}^{\rm dr}\widehat u_{\gtrless}$.

	The half-line monodromy relation reads
	\begin{equation}
		\mathcal T_{\rm f}\Omega=\mathcal T_{\rm i},\qquad
		\Omega\equiv G_{\rm f}(0)^{-1}G_{\rm i}(0).
		\label{eq:Trelation-main}
	\end{equation}
    To obtain the endpoint relation, we first consider the transfer matrix on a
    finite interval. The zero-curvature condition shows that the time derivative
    of $\mathcal T_{(L,0)}(t)G(0,t)^{-1}$ is controlled only by the temporal Lax
    matrix at $s=L$. For the physical MFT saddle, parabolic smoothing makes the
    scattering fields $u$ and $v$ exponentially localized in $s$. The boundary
    term therefore vanishes as $L\to\infty$, which gives
    $\partial_t[\mathcal T(t)G(0,t)^{-1}]=0$. Comparing the two temporal
    endpoints then yields Eq.~\eqref{eq:Trelation-main}. This relation is first established in
    a regular neighborhood of $z=1$, where $k(t;1)=0$, and is then extended by
    analytic continuation to the common domain of analyticity of both sides
    \cite{SM}.
    
	Writing $\eta=(\Omega)_{11}$, the fixed endpoint frames give
	$\mathsf{u}\mathsf{v}=\eta\omega$.  Comparing the appropriate matrix entries then
	yields the exterior-channel identity
	\begin{equation}
		\mathsf V_>(z)\mathsf U_>(z)
		=\eta[1+\omega K(z)],
		\label{eq:VU-factor-main}
	\end{equation}
		where $K(z)=\exp\!\left[\frac{r_0^2}{4}
		\left(z+z^{-1}-2\right)\right]$. 
		The canonical spectral contour is the unit circle $|z|=1$. 
		The subsequent scalar factorization is a Riemann--Hilbert (equivalently,
		Wiener--Hopf) factorization on this contour~\cite{BealsCoifman1984,Fokas2008}.
		A complementary comparison of the same matrix identity reconstructs
		$\mathsf V_>$ from the finite-interval channel $\mathsf U_<$ and
		$\mathsf U_>$ from $\mathsf V_<$.  These finite-interval transforms fix
		the separate analytic assignment: $\mathsf V_>(0)=1$ and
		$\mathsf U_>(\infty)=1$, with $\mathsf V_>$ analytic in
		$|z|<1$ and $\mathsf U_>$ analytic in $|z|>1$; in particular,
		neither poles nor zeros occur in these respective domains. 
	Let $\ell(\theta)= \log (1+\omega K(e^{\ii\theta}))$ and
	$\ell_n=(2\pi)^{-1}\int_0^{2\pi}\ell(\theta)e^{-\ii n\theta}\dd\theta$,
	then
	\begin{equation}
		\log\eta=-\ell_0,~~
		\mathsf V_>(z)=\exp\!\bigg(\sum_{n\ge1}\ell_n z^n\bigg),\quad |z|<1,
		\label{eq:Vfactor-main}
	\end{equation}
	and $\mathsf U_>(z)=\mathsf V_>(z^{-1})$ for $|z|>1$. Using the singular
	interface terms in $u,v$ and the far-field condition fixes the shear
	amplitudes; in particular,
	$\mathsf{u}
	=
	\mathsf V_{>}(1)\frac{
		\rho_1(1-\rho_2)e^\lambda
		-\rho_2(1-\rho_1)
	}
	{
		1-\rho_1+\rho_1e^\lambda
	}$. 
	The fixed-frame evaluation of $\mathsf{u},\mathsf{v},\eta$ and the detailed
	analytic-continuation argument are given in \cite{SM}.
	
	\begin{figure}[!t]
		\centering
		\includegraphics[width=0.95\columnwidth]{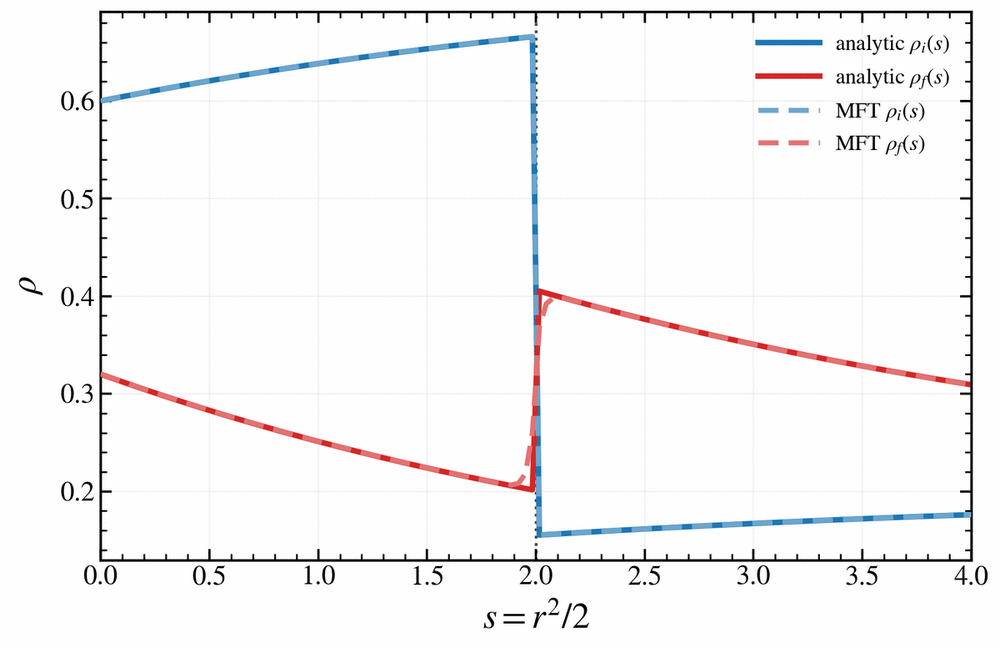}
		\caption{Analytic endpoint-density reconstruction versus the direct
			iterative MFT solution for $r_0=2$ and
			$(\rho_1,\rho_2,\lambda)=(0.5,0.2,1)$.  In the notation of \cite{SM},
			the curves show
			$\rho_{\rm i}(s)$ and $\rho_{\rm f}(s)$; solid lines are analytic and dashed lines
			are numerical.}
		\label{fig:reconstruction}
	\end{figure}
	
	\Sec{Spectral Reconstruction of the SCGF}
	The zero of the wave number at $z=1$ makes this point the natural place to
	extract the transported mass.  Define $\mathcal V_>^{(1)}$ by the local
	expansion
	$\log\mathsf V_>(z)=\log\mathsf V_>(1)
	+\mathcal V_>^{(1)}(z-1)+O((z-1)^2)$.
	Eq.~\eqref{eq:Vfactor-main} gives
	$\mathcal V_>^{(1)}=\sum_{n\geq1}n\ell_n$.  On the other hand, expanding
	the exterior Jost solutions at the same point expresses this coefficient in
	terms of the excess masses outside the disk at the two temporal endpoints.
	The endpoint jump relations then give the zero-mode trace identity
	\begin{equation}
		q(\lambda)=4\pi\mathcal V_>^{(1)}
		\frac{\omega'(\lambda)}{\omega(\lambda)}.
		\label{eq:q-trace-main}
		\end{equation}
		At a zero of $\omega$, the right-hand side of
		Eq.~\eqref{eq:q-trace-main} is understood by its removable continuous
		extension: since $\ell_n=O(\omega)$, one has
		$\mathcal V_>^{(1)}=O(\omega)$.
		Since $q=\partial_\lambda\mu_2^{\rm ann}
	=\omega'\partial_\omega\mathsf F_2$, Eq.~\eqref{eq:q-trace-main} yields
	\begin{equation}
		\omega\,\partial_\omega \mathsf{F}_2(\omega)
		=4\pi \mathcal{V}_>^{(1)}.
		\label{eq:trace-main}
	\end{equation}
	This relation also shows explicitly why the first coefficient at the regular
	spectral point, rather than a large-$k$ coefficient, generates the current in
	the nonisospectral problem.
	To evaluate Eq.~\eqref{eq:trace-main}, use
	$\ell(0)-\ell(\theta)=2\sum_{n\geq1}\ell_n(1-\cos n\theta)$ and
	$\int_0^{2\pi}(1-\cos n\theta)/(1-\cos\theta)\,\dd\theta=2\pi n$.  Integrating by parts then gives
	\begin{equation}
		\partial_\omega\mathsf F_2
		=2r_0^2\int_0^\pi\!\cos^2\frac{\theta}{2}
		\frac{e^{-r_0^2\sin^2(\theta/2)}}
		{1+\omega e^{-r_0^2\sin^2(\theta/2)}}\,\dd\theta.
		\label{eq:dF-main}
	\end{equation}
	Integrating from $0$ to $\omega$ and imposing $\mathsf F_2(0)=0$ gives
	Eq.~\eqref{eq:mainresult}.  Thus the scalar factorization, its zero-mode
	coefficient, and the SCGF are linked without an undetermined integration
	constant.

	\Sec{Explicit Cumulants and Independent Checks}
	Expanding the exact result as
	$\mathsf F_2(\omega)=\sum_{n\geq1}(-1)^{n+1}m_n\omega^n/n$
	defines geometric channel moments.  The angular integral can be performed in
	terms of modified Bessel functions:
	\begin{equation}
		m_n=\pi r_0^2e^{-nr_0^2/2}
		\left[I_0\!\left(\frac{nr_0^2}{2}\right)
		+I_1\!\left(\frac{nr_0^2}{2}\right)\right].
		\label{eq:mn-main}
	\end{equation}
	A direct perturbative solution of the MFT equations to cubic order gives
	Eq.~\eqref{eq:mn-main} independently for $n=1,2,3$; the heat-kernel
	contractions and response hierarchy are presented in \cite{SM}.  These checks probe the result before any use of the scalar
	factorization.  Figure~\ref{fig:scgf-rate}(b) shows that higher cumulants remain visible
	beyond the Gaussian neighborhood of the typical current.

	\Sec{Geometric Limits}
	The exact finite-radius result interpolates between two geometrically distinct
	regimes.  For a small disk, its diffusive relaxation time $O(r_0^2)$ is short
	compared with the observation time.  The initial and final endpoint tilts then
	decouple to leading order, giving
	\begin{equation}
		\mu_2^{\rm ann}(\lambda;r_0)
		=\pi r_0^2\log[1+\omega(\lambda)]+o(r_0^2),
		\quad r_0\to0.
		\label{eq:small-r-main}
	\end{equation}
	The prefactor is the disk area, and the logarithm is the generating function
	of the difference of the two independently tilted Bernoulli endpoint
	occupations.  Thus the small-radius regime is controlled by the counting
	volume rather than by its boundary.

	For $r_0\gg1$, the optimal fields are confined to an $O(1)$ diffusive layer
	around the circle.  Curvature is subleading inside this layer, so the saddle is
	locally the planar one-interface SSEP saddle.  Eq.~\eqref{eq:mainresult} gives
	\begin{equation}
		\mu_2^{\rm ann}(\lambda;r_0)
		=-2\sqrt{\pi}\,r_0
		\Li_{3/2}[-\omega(\lambda)]+o(r_0),
		\quad r_0\to\infty.
		\label{eq:large-r-main}
	\end{equation}
	Here the leading prefactor is proportional to the circumference, and the
	remaining function is precisely the universal planar SCGF per unit interface
	length.  The corresponding arbitrary-dimensional limits are given in \cite{SM}.

    \Sec{Optimal-Density Reconstruction}
    The scalar factor contains more information than the integrated current: it
    also reconstructs the optimal density profiles at the initial and final
    times. Fourier inversion of the scalar factor gives the cumulative scattering
    potentials
    \begin{align}
    	\mathcal G_{>}(\ell)
    	&:=
    	\mathsf{u}\int_\ell^\infty v_{\rm i}(s_0+q)\dd q,
    	&& \ell>0, \label{eq:Gout-main}\\
    	\mathcal G_{<}(\ell)
    	&:=
    	\mathsf{v}\int_0^\ell u_{\rm f}(s_0-q)\dd q,
    	&& 0<\ell<s_0. \label{eq:Gin-main}
    \end{align}
    The endpoint duality $\mathsf{u}v_{\rm i}(s)=\mathsf{v}u_{\rm f}(s)$ away
    from the counting interface shows that either endpoint may equivalently be
    used to define these potentials.
    
    Since $v_{\rm i}(s)$ and $u_{\rm f}(s)$ are proportional to
    $\partial_s\rho_{\rm i}(s)$ and $\partial_s\rho_{\rm f}(s)$ respectively,  a single spatial integration reconstructs the profiles. For
    example, outside the disk,
    \begin{align}
    	\rho_{\rm i}(s_0+\ell)
    	&=\rho_2-\frac{\rho_2(1-\rho_2)}{\mathsf{u}}\,
    	\mathcal G_{>}(\ell), \label{eq:rhoi-reconstruction-main}\\
    	\rho_{\rm f}(s_0+\ell)
    	&=\rho_2+\frac{1}{\mathsf{v}}\,
    	\mathcal G_{>}(\ell). \label{eq:rhof-reconstruction-main}
    \end{align}
    The corresponding interior formulas are expressed through
    $\mathcal G_{<}$. The interface values follow from the endpoint jump
    conditions, while origin regularity and the far-field density fix the
    remaining constants. Explicit stable inversion and interior reconstruction
    formulas are given in \cite{SM}. The parameter-free agreement in
    Fig.~\ref{fig:reconstruction} tests the scattering solution at the field
    level.

	\Sec{Spectral Channel Representation}
	The exact solution also admits the channel form
	\begin{equation}
		\mathsf{F}_2(\omega)=\int_0^1\log(1+\omega\theta)\, \nu_2(\dd\theta),
		\label{eq:channel-rep}
	\end{equation}
	with
	\begin{equation}
		\frac{\dd \nu_2}{\dd\theta}=\frac{2\sqrt{r_0^2+\log\theta}}{\theta\sqrt{-\log\theta}}\,
		\mathbf 1_{\{e^{-r_0^2}<\theta<1\}}.
		\label{eq:channel}
	\end{equation}
	The full counting statistics can therefore be written as a continuum of
	effective Bernoulli channels, with the radial geometry encoded in $\nu_2$.
	This channel structure is the continuum counterpart of one-dimensional SSEP
	representations \cite{DerridaGerschenfeld2009JSP136,DerridaGerschenfeld2009JSP137}.
	The measure is positive, supported on $e^{-r_0^2}<\theta<1$, and has total
	weight $\nu_2([0,1])=\pi r_0^2$; its moments satisfy
	$\int_0^1\theta^n\nu_2(\dd\theta)=m_n$, with $m_n$ from
	Eq.~\eqref{eq:mn-main}.  Hence $\mathsf F_2$ is increasing and concave for
	$\omega>-1$, while the lower edge of the support moving toward zero as
	$r_0$ grows yields the planar limit.
	
	\Sec{Discussion and Outlook}
	The finite-radius solution is exceptional to the radial two-dimensional
	problem.  For $d\ge3$, the radial volume coordinate $s=r^d/d$ flattens the
	measure but leaves a position-dependent coefficient, so the single-parameter
	Lax pair and scalar factorization do not close; the arbitrary-dimensional
	limits are therefore asymptotic results \cite{SM}.  In $d=1$, the standard
	step state on the full line has a known exact SCGF
	\cite{DerridaGerschenfeld2009JSP136,DerridaGerschenfeld2009JSP137,
	MallickMoriyaSasamoto2022,MallickMoriyaSasamoto2024}, whereas a
	reflection-symmetric finite interval requires additional reflection data and
	is not obtained by directly specializing the present factorization.  This is
	an additional inverse-scattering problem, not a failure of bulk integrability.

	The continuum channel measure nevertheless suggests a useful microscopic
	question.  Strong-Rayleigh theory gives finite-system Poisson-binomial
	factorizations for SSEP occupation generating polynomials on general graphs
	\cite{BorceaBrandenLiggett2009}.  It would be interesting to determine whether
	a finite-system spectral measure for the disk converges to $\nu_2$ in the
	diffusive limit.  Quenched initial conditions and other diffusive lattice gases
	are further natural tests of which parts of the channel picture survive beyond
	the annealed two-dimensional setting.

	A further direction is to replace the scalar particle density by a
	vector-valued conserved field, or by several noncommuting conserved charges,
	while retaining a finite two-dimensional counting region.  This would combine
	curved two-dimensional MFT with genuinely non-Abelian quantum transport.  Recent
	work derives fluctuation relations and uncertainty bounds for the exchange
	statistics of multiple noncommuting conserved charges
	\cite{ScandiManzano2026}; adapting these ideas to a spatially extended,
	vector-valued current large-deviation problem is left for future work.
	
	\Sec{Acknowledgments}
	OpenAI ChatGPT 5.6 was used as an AI-assisted tool for portions of the numerical implementation, including the Monte Carlo simulations; for preparing the schematic illustration in Fig.~\ref{fig:model} and other numerical plots; and for assisting with parts of the analytical derivations and manuscript preparation. The author independently checked and verified all code, figures, derivations, equations, and numerical results and takes full responsibility for the content. Tingfei Li acknowledges support from the National Natural Science Foundation of China (Grant No.~12605129). The author also thanks Xingpao Suo for valuable discussions.
	
	\Sec{Data Availability}
	The numerical endpoint-density data shown in Fig.~\ref{fig:reconstruction} were generated for this work from the MFT equations and the parameter set specified in the caption; no external data sets were used.
	
\bibliographystyle{apsrev4-2}
\bibliography{refer2}

@article{Spitzer1970,
  author = {Frank Spitzer},
  title = {Interaction of Markov processes},
  journal = {Adv. Math.},
  volume = {5},
  pages = {246--290},
  year = {1970},
  doi = {10.1016/0001-8708(70)90034-4}
}

@book{Liggett1985,
  author = {Thomas M. Liggett},
  title = {Interacting Particle Systems},
  publisher = {Springer},
  address = {New York},
  year = {1985},
  doi = {10.1007/978-1-4613-8542-4}
}

@book{Spohn1991,
  author = {Herbert Spohn},
  title = {Large Scale Dynamics of Interacting Particles},
  publisher = {Springer},
  address = {Berlin},
  year = {1991},
  doi = {10.1007/978-3-642-84371-6}
}

@book{KipnisLandim1999,
  author = {Claude Kipnis and Claudio Landim},
  title = {Scaling Limits of Interacting Particle Systems},
  publisher = {Springer},
  address = {Berlin},
  series = {Grundlehren der mathematischen Wissenschaften},
  volume = {320},
  year = {1999},
  doi = {10.1007/978-3-662-03752-2}
}

@book{DemboZeitouni1998,
  author = {Amir Dembo and Ofer Zeitouni},
  title = {Large Deviations Techniques and Applications},
  publisher = {Springer},
  address = {New York},
  edition = {2},
  year = {1998},
  doi = {10.1007/978-1-4612-5320-4}
}

@article{Jensen1906,
  author = {Johan Ludwig William Valdemar Jensen},
  title = {Sur les fonctions convexes et les in{\'e}galit{\'e}s entre les valeurs moyennes},
  journal = {Acta Mathematica},
  volume = {30},
  pages = {175--193},
  year = {1906},
  doi = {10.1007/BF02418571}
}

@book{FreidlinWentzell2012,
  author = {Mark I. Freidlin and Alexander D. Wentzell},
  title = {Random Perturbations of Dynamical Systems},
  publisher = {Springer},
  address = {Berlin},
  edition = {3},
  year = {2012},
  doi = {10.1007/978-3-642-25847-3}
}

@article{Touchette2009,
  author = {Hugo Touchette},
  title = {The large deviation approach to statistical mechanics},
  journal = {Phys. Rep.},
  volume = {478},
  pages = {1--69},
  year = {2009},
  doi = {10.1016/j.physrep.2009.05.002}
}

@article{GallavottiCohen1995,
  author = {Giovanni Gallavotti and E. G. D. Cohen},
  title = {Dynamical ensembles in nonequilibrium statistical mechanics},
  journal = {Phys. Rev. Lett.},
  volume = {74},
  pages = {2694--2697},
  year = {1995},
  doi = {10.1103/PhysRevLett.74.2694}
}

@article{LebowitzSpohn1999,
  author = {Joel L. Lebowitz and Herbert Spohn},
  title = {A Gallavotti--Cohen-type symmetry in the large deviation functional for stochastic dynamics},
  journal = {J. Stat. Phys.},
  volume = {95},
  pages = {333--365},
  year = {1999},
  doi = {10.1023/A:1004589714161}
}

@article{Bertini2002,
  author = {Lorenzo Bertini and Alberto De Sole and Davide Gabrielli and Giovanni Jona-Lasinio and Claudio Landim},
  title = {Macroscopic fluctuation theory for stationary non-equilibrium states},
  journal = {J. Stat. Phys.},
  volume = {107},
  pages = {635--675},
  year = {2002},
  doi = {10.1023/A:1014525911391}
}

@article{Bertini2003,
  author = {Lorenzo Bertini and Alberto De Sole and Davide Gabrielli and Giovanni Jona-Lasinio and Claudio Landim},
  title = {Large deviations for the boundary driven symmetric simple exclusion process},
  journal = {Math. Phys. Anal. Geom.},
  volume = {6},
  pages = {231--267},
  year = {2003},
  doi = {10.1023/A:1024967818899}
}

@article{Bertini2005PRL,
  author = {Lorenzo Bertini and Alberto De Sole and Davide Gabrielli and Giovanni Jona-Lasinio and Claudio Landim},
  title = {Current fluctuations in stochastic lattice gases},
  journal = {Phys. Rev. Lett.},
  volume = {94},
  pages = {030601},
  year = {2005},
  doi = {10.1103/PhysRevLett.94.030601}
}

@article{Bertini2006JSP,
  author = {Lorenzo Bertini and Alberto De Sole and Davide Gabrielli and Giovanni Jona-Lasinio and Claudio Landim},
  title = {Non-equilibrium current fluctuations in stochastic lattice gases},
  journal = {J. Stat. Phys.},
  volume = {123},
  pages = {237--276},
  year = {2006},
  doi = {10.1007/s10955-006-9056-4}
}

@article{BertiniHeat2005,
  author = {Lorenzo Bertini and Davide Gabrielli and Joel L. Lebowitz},
  title = {Large deviations for a stochastic model of heat flow},
  journal = {J. Stat. Phys.},
  volume = {121},
  pages = {843--885},
  year = {2005},
  doi = {10.1007/s10955-005-5527-2}
}

@article{BertiniRMP2015,
  author = {Lorenzo Bertini and Alberto De Sole and Davide Gabrielli and Giovanni Jona-Lasinio and Claudio Landim},
  title = {Macroscopic fluctuation theory},
  journal = {Rev. Mod. Phys.},
  volume = {87},
  pages = {593--636},
  year = {2015},
  doi = {10.1103/RevModPhys.87.593}
}

@article{DerridaReview2007,
  author = {Bernard Derrida},
  title = {Non-equilibrium steady states: fluctuations and large deviations of the density and of the current},
  journal = {J. Stat. Mech.},
  pages = {P07023},
  year = {2007},
  doi = {10.1088/1742-5468/2007/07/P07023}
}

@article{DerridaLebowitzSpeer2002JSP,
  author = {Bernard Derrida and Joel L. Lebowitz and Eugene R. Speer},
  title = {Large deviation of the density profile in the steady state of the open symmetric simple exclusion process},
  journal = {J. Stat. Phys.},
  volume = {107},
  pages = {599--634},
  year = {2002},
  doi = {10.1023/A:1014555927320}
}

@article{DerridaLebowitzSpeer2002PRL,
  author = {Bernard Derrida and Joel L. Lebowitz and Eugene R. Speer},
  title = {Exact free energy functional for a driven diffusive open stationary nonequilibrium system},
  journal = {Phys. Rev. Lett.},
  volume = {89},
  pages = {030601},
  year = {2002},
  doi = {10.1103/PhysRevLett.89.030601}
}

@article{DerridaLebowitzSpeer2003,
  author = {Bernard Derrida and Joel L. Lebowitz and Eugene R. Speer},
  title = {Exact large deviation functional of a stationary open driven diffusive system: the asymmetric exclusion process},
  journal = {J. Stat. Phys.},
  volume = {110},
  pages = {775--810},
  year = {2003},
  doi = {10.1023/A:1022111919402}
}

@article{EnaudDerrida2004,
  author = {C. Enaud and Bernard Derrida},
  title = {Large deviation functional of the weakly asymmetric exclusion process},
  journal = {J. Stat. Phys.},
  volume = {114},
  pages = {537--562},
  year = {2004},
  doi = {10.1023/B:JOSS.0000012501.43746.CF}
}

@article{BodineauDerrida2004,
  author = {Thierry Bodineau and Bernard Derrida},
  title = {Current fluctuations in nonequilibrium diffusive systems: An additivity principle},
  journal = {Phys. Rev. Lett.},
  volume = {92},
  pages = {180601},
  year = {2004},
  doi = {10.1103/PhysRevLett.92.180601}
}

@article{BodineauDerrida2005,
  author = {Thierry Bodineau and Bernard Derrida},
  title = {Distribution of current in nonequilibrium diffusive systems and phase transitions},
  journal = {Phys. Rev. E},
  volume = {72},
  pages = {066110},
  year = {2005},
  doi = {10.1103/PhysRevE.72.066110}
}

@article{BodineauDerrida2007,
  author = {Thierry Bodineau and Bernard Derrida},
  title = {Cumulants and large deviations of the current through non-equilibrium steady states},
  journal = {C. R. Physique},
  volume = {8},
  pages = {540--555},
  year = {2007},
  doi = {10.1016/j.crhy.2007.04.014}
}

@article{LecomteEtAl2007,
  author = {Vivien Lecomte and C\'ecile Appert-Rolland and Fr\'ed\'eric van Wijland},
  title = {Thermodynamic formalism for systems with Markov dynamics},
  journal = {J. Stat. Phys.},
  volume = {127},
  pages = {51--106},
  year = {2007},
  doi = {10.1007/s10955-006-9254-0}
}

@article{AppertRollandEtAl2008,
  author = {C\'ecile Appert-Rolland and Bernard Derrida and Vivien Lecomte and Fr\'ed\'eric van Wijland},
  title = {Universal cumulants of the current in diffusive systems on a ring},
  journal = {Phys. Rev. E},
  volume = {78},
  pages = {021122},
  year = {2008},
  doi = {10.1103/PhysRevE.78.021122}
}

@article{HurtadoGarrido2009,
  author = {Pablo I. Hurtado and Pedro L. Garrido},
  title = {Test of the additivity principle for current fluctuations in a model of heat conduction},
  journal = {Phys. Rev. Lett.},
  volume = {102},
  pages = {250601},
  year = {2009},
  doi = {10.1103/PhysRevLett.102.250601}
}

@article{HurtadoGarrido2011,
  author = {Pablo I. Hurtado and Pedro L. Garrido},
  title = {Spontaneous symmetry breaking at the fluctuating level},
  journal = {Phys. Rev. Lett.},
  volume = {107},
  pages = {180601},
  year = {2011},
  doi = {10.1103/PhysRevLett.107.180601}
}

@article{DerridaDoucotRoche2004,
  author = {Bernard Derrida and Brigitte Dou\c{c}ot and Pierre-Etienne Roche},
  title = {Current fluctuations in the one-dimensional symmetric exclusion process with open boundaries},
  journal = {J. Stat. Phys.},
  volume = {115},
  pages = {717--748},
  year = {2004},
  doi = {10.1023/B:JOSS.0000022379.95508.b2}
}

@article{DerridaGerschenfeld2009JSP136,
  author = {Bernard Derrida and Antoine Gerschenfeld},
  title = {Current fluctuations of the one dimensional symmetric simple exclusion process with step initial condition},
  journal = {J. Stat. Phys.},
  volume = {136},
  pages = {1--15},
  year = {2009},
  doi = {10.1007/s10955-009-9772-7}
}

@article{DerridaGerschenfeld2009JSP137,
  author = {Bernard Derrida and Antoine Gerschenfeld},
  title = {Current fluctuations in one dimensional diffusive systems with a step initial density profile},
  journal = {J. Stat. Phys.},
  volume = {137},
  pages = {978--1000},
  year = {2009},
  doi = {10.1007/s10955-009-9830-1}
}

@article{DeMasiFerrari2002,
  author = {Anna De Masi and Pablo A. Ferrari},
  title = {Flux fluctuations in the one dimensional nearest neighbors symmetric simple exclusion process},
  journal = {J. Stat. Phys.},
  volume = {107},
  pages = {677--683},
  year = {2002},
  doi = {10.1023/A:1014577928229}
}

@article{VandenbergRodes2010,
  author = {Alexander Vandenberg-Rodes},
  title = {A limit theorem for particle current in the symmetric exclusion process},
  journal = {Electron. Commun. Probab.},
  volume = {15},
  pages = {240--252},
  year = {2010},
  doi = {10.1214/ECP.v15-1550}
}

@article{SethuramanVaradhan2013,
  author = {Sunder Sethuraman and S. R. S. Varadhan},
  title = {Large deviations for the current and tagged particle in 1D nearest-neighbor symmetric simple exclusion},
  journal = {Ann. Probab.},
  volume = {41},
  pages = {1461--1512},
  year = {2013},
  doi = {10.1214/11-AOP703}
}

@article{Arratia1983,
  author = {Richard Arratia},
  title = {The motion of a tagged particle in the simple symmetric exclusion system on $\mathbb{Z}$},
  journal = {Ann. Probab.},
  volume = {11},
  pages = {362--373},
  year = {1983},
  doi = {10.1214/aop/1176993602}
}

@article{KrapivskyMeerson2012,
  author = {P. L. Krapivsky and Baruch Meerson},
  title = {Fluctuations of current in nonstationary diffusive lattice gases},
  journal = {Phys. Rev. E},
  volume = {86},
  pages = {031106},
  year = {2012},
  doi = {10.1103/PhysRevE.86.031106}
}

@article{VilenkinMeersonSasorov2014,
  author = {Arkady Vilenkin and Baruch Meerson and Pavel V. Sasorov},
  title = {Extreme fluctuations of current in the symmetric simple exclusion process: A non-stationary setting},
  journal = {J. Stat. Mech.},
  pages = {P06007},
  year = {2014},
  doi = {10.1088/1742-5468/2014/06/P06007}
}

@article{KrapivskyMallickSadhu2014,
  author = {P. L. Krapivsky and Kirone Mallick and Tridib Sadhu},
  title = {Large deviations in single-file diffusion},
  journal = {Phys. Rev. Lett.},
  volume = {113},
  pages = {078101},
  year = {2014},
  doi = {10.1103/PhysRevLett.113.078101}
}

@article{KrapivskyMallickSadhu2015,
  author = {P. L. Krapivsky and Kirone Mallick and Tridib Sadhu},
  title = {Tagged particle in single-file diffusion},
  journal = {J. Stat. Phys.},
  volume = {160},
  pages = {885--925},
  year = {2015},
  doi = {10.1007/s10955-015-1291-0}
}

@article{KrapivskyMallickSadhu2015Dyn,
  author = {P. L. Krapivsky and Kirone Mallick and Tridib Sadhu},
  title = {Dynamical properties of single-file diffusion},
  journal = {J. Stat. Mech.},
  pages = {P09007},
  year = {2015},
  doi = {10.1088/1742-5468/2015/09/P09007}
}

@article{SadhuDerrida2015,
  author = {Tridib Sadhu and Bernard Derrida},
  title = {Large deviation function of a tracer position in single file diffusion},
  journal = {J. Stat. Mech.},
  pages = {P09008},
  year = {2015},
  doi = {10.1088/1742-5468/2015/09/P09008}
}

@article{ImamuraMallickSasamoto2021,
  author = {Takashi Imamura and Kirone Mallick and Tomohiro Sasamoto},
  title = {Distribution of a tagged particle position in the one-dimensional symmetric simple exclusion process with two-sided Bernoulli initial condition},
  journal = {Commun. Math. Phys.},
  volume = {384},
  pages = {1409--1444},
  year = {2021},
  doi = {10.1007/s00220-021-03954-x}
}

@article{GrabschEtAl2022,
  author = {Aur\'elien Grabsch and Alexis Poncet and Pierre Rizkallah and Pierre Illien and Olivier B\'enichou},
  title = {Exact closure and solution for spatial correlations in single-file diffusion},
  journal = {Sci. Adv.},
  volume = {8},
  pages = {eabm5043},
  year = {2022},
  doi = {10.1126/sciadv.abm5043}
}

@article{MallickMoriyaSasamoto2022,
  author = {Kirone Mallick and Hiroki Moriya and Tomohiro Sasamoto},
  title = {Exact solution of the macroscopic fluctuation theory for the symmetric exclusion process},
  journal = {Phys. Rev. Lett.},
  volume = {129},
  pages = {040601},
  year = {2022},
  doi = {10.1103/PhysRevLett.129.040601}
}

@article{MallickMoriyaSasamoto2024,
  author = {Kirone Mallick and Hiroki Moriya and Tomohiro Sasamoto},
  title = {Exact solutions to macroscopic fluctuation theory through classical integrable systems},
  journal = {J. Stat. Mech.},
  volume = {2024},
  pages = {074001},
  year = {2024},
  doi = {10.1088/1742-5468/ad485e}
}

@article{GrabschEtAl2024,
  author = {Aur\'elien Grabsch and Hiroki Moriya and Kirone Mallick and Tomohiro Sasamoto and Olivier B\'enichou},
  title = {Semi-infinite simple exclusion process: From current fluctuations to target survival},
  journal = {Phys. Rev. Lett.},
  volume = {133},
  pages = {117102},
  year = {2024},
  doi = {10.1103/PhysRevLett.133.117102}
}

@article{BerliozEtAl2024,
  author = {Th\'eotim Berlioz and Davide Venturelli and Aur\'elien Grabsch and Olivier B\'enichou},
  title = {Current fluctuations in the symmetric exclusion process beyond the one-dimensional geometry},
  journal = {J. Stat. Mech.},
  volume = {2024},
  pages = {113208},
  year = {2024},
  doi = {10.1088/1742-5468/ad874a}
}

@article{SahaEtAl2026,
  author = {Kapil Sharma and Soumyabrata Saha and Sandeep Jangid and Tridib Sadhu},
  title = {Large deviations of current for the symmetric simple exclusion process on a semi-infinite line and on an infinite line with a slow bond},
  journal = {Phys. Rev. E},
  volume = {113},
  pages = {L052101},
  year = {2026},
  doi = {10.1103/dzf3-8wpt}
}

@article{AkkermansEtAl2013,
  author = {Eric Akkermans and Thierry Bodineau and Bernard Derrida and Ohad Shpielberg},
  title = {Universal current fluctuations in the symmetric exclusion process and other diffusive systems},
  journal = {EPL},
  volume = {103},
  pages = {20001},
  year = {2013},
  doi = {10.1209/0295-5075/103/20001}
}

@article{MeersonVilenkinKrapivsky2014,
  author = {Baruch Meerson and Arkady Vilenkin and P. L. Krapivsky},
  title = {Survival of a static target in a gas of diffusing particles with exclusion},
  journal = {Phys. Rev. E},
  volume = {90},
  pages = {022120},
  year = {2014},
  doi = {10.1103/PhysRevE.90.022120}
}

@article{Meerson2015Absorption,
  author = {Baruch Meerson},
  title = {Full absorption statistics of diffusing particles with exclusion},
  journal = {J. Stat. Mech.},
  pages = {P04009},
  year = {2015},
  doi = {10.1088/1742-5468/2015/04/P04009}
}

@article{AgranovMeerson2017,
  author = {Tal Agranov and Baruch Meerson},
  title = {Fluctuations of absorption of interacting diffusing particles by multiple absorbers},
  journal = {Phys. Rev. E},
  volume = {95},
  pages = {062124},
  year = {2017},
  doi = {10.1103/PhysRevE.95.062124}
}

@article{TracyWidom2008,
  author = {Craig A. Tracy and Harold Widom},
  title = {Integral formulas for the asymmetric simple exclusion process},
  journal = {Commun. Math. Phys.},
  volume = {279},
  pages = {815--844},
  year = {2008},
  doi = {10.1007/s00220-008-0443-3}
}

@article{TracyWidom2009,
  author = {Craig A. Tracy and Harold Widom},
  title = {Asymptotics in ASEP with step initial condition},
  journal = {Commun. Math. Phys.},
  volume = {290},
  pages = {129--154},
  year = {2009},
  doi = {10.1007/s00220-009-0761-0}
}

@article{TailleurKurchanLecomte2008,
  author = {Julien Tailleur and Jorge Kurchan and Vivien Lecomte},
  title = {Mapping out-of-equilibrium into equilibrium in one-dimensional transport models},
  journal = {J. Phys. A},
  volume = {41},
  pages = {505001},
  year = {2008},
  doi = {10.1088/1751-8113/41/50/505001}
}

@article{GorissenVanderzande2012,
  author = {Mieke Gorissen and Carlo Vanderzande},
  title = {Current fluctuations in the weakly asymmetric exclusion process with open boundaries},
  journal = {Phys. Rev. E},
  volume = {86},
  pages = {051114},
  year = {2012},
  doi = {10.1103/PhysRevE.86.051114}
}

@article{XueZhao2024,
  author = {Xiaofeng Xue and Linjie Zhao},
  title = {Moderate deviations for the current and tagged particle in symmetric simple exclusion processes},
  journal = {Stochastic Process. Appl.},
  volume = {167},
  pages = {104218},
  year = {2024},
  doi = {10.1016/j.spa.2023.09.005}
}

@article{KipnisMarchioroPresutti1982,
  author = {Claude Kipnis and Claudio Marchioro and Errico Presutti},
  title = {Heat flow in an exactly solvable model},
  journal = {J. Stat. Phys.},
  volume = {27},
  pages = {65--74},
  year = {1982},
  doi = {10.1007/BF01011740}
}

@article{BettelheimSmithMeerson2022PRL,
  author = {Eldad Bettelheim and Naftali R. Smith and Baruch Meerson},
  title = {Inverse scattering method solves the problem of full statistics of nonstationary heat transfer in the Kipnis--Marchioro--Presutti model},
  journal = {Phys. Rev. Lett.},
  volume = {128},
  pages = {130602},
  year = {2022},
  doi = {10.1103/PhysRevLett.128.130602}
}

@article{BettelheimSmithMeerson2022JSTAT,
  author = {Eldad Bettelheim and Naftali R. Smith and Baruch Meerson},
  title = {Full statistics of nonstationary heat transfer in the Kipnis--Marchioro--Presutti model},
  journal = {J. Stat. Mech.},
  volume = {2022},
  pages = {093103},
  year = {2022},
  doi = {10.1088/1742-5468/ac8a4d}
}

@article{Bettelheim2024Whitham,
  author = {Eldad Bettelheim},
  title = {Whitham approach to certain large fluctuation problems in statistical mechanics},
  journal = {J. Phys. A},
  volume = {57},
  pages = {035201},
  year = {2024},
  doi = {10.1088/1751-8121/ad17d6}
}

@article{BettelheimMeerson2024,
  author = {Eldad Bettelheim and Baruch Meerson},
  title = {Complete integrability of the problem of full statistics of nonstationary mass transfer in the simple inclusion process},
  journal = {Phys. Rev. E},
  volume = {110},
  pages = {014101},
  year = {2024},
  doi = {10.1103/PhysRevE.110.014101}
}

@article{Tsai2023,
  author = {Li-Cheng Tsai},
  title = {Integrability in the weak noise theory},
  journal = {Trans. Am. Math. Soc.},
  volume = {376},
  pages = {6521--6572},
  year = {2023},
  doi = {10.1090/tran/8977}
}

@article{KrajenbrinkLeDoussal2021,
  author = {Alexandre Krajenbrink and Pierre Le Doussal},
  title = {Inverse scattering of the Zakharov--Shabat system solves the weak noise theory of the Kardar--Parisi--Zhang equation},
  journal = {Phys. Rev. Lett.},
  volume = {127},
  pages = {064101},
  year = {2021},
  doi = {10.1103/PhysRevLett.127.064101}
}

@article{KrajenbrinkLeDoussal2022,
  author = {Alexandre Krajenbrink and Pierre Le Doussal},
  title = {Inverse scattering solution of the weak noise theory of the Kardar--Parisi--Zhang equation with flat and Brownian initial conditions},
  journal = {Phys. Rev. E},
  volume = {105},
  pages = {054142},
  year = {2022},
  doi = {10.1103/PhysRevE.105.054142}
}

@article{KrajenbrinkLeDoussal2023,
  author = {Alexandre Krajenbrink and Pierre Le Doussal},
  title = {Crossover from the macroscopic fluctuation theory to the Kardar--Parisi--Zhang equation controls the large deviations beyond Einstein's diffusion},
  journal = {Phys. Rev. E},
  volume = {107},
  pages = {014137},
  year = {2023},
  doi = {10.1103/PhysRevE.107.014137}
}

@article{JanasKamenevMeerson2016,
  author = {M. Janas and Alex Kamenev and Baruch Meerson},
  title = {Dynamical phase transition in large-deviation statistics of the Kardar--Parisi--Zhang equation},
  journal = {Phys. Rev. E},
  volume = {94},
  pages = {032133},
  year = {2016},
  doi = {10.1103/PhysRevE.94.032133}
}

@article{AKNS1974,
  author = {Mark J. Ablowitz and David J. Kaup and Alan C. Newell and Harvey Segur},
  title = {The inverse scattering transform---Fourier analysis for nonlinear problems},
  journal = {Stud. Appl. Math.},
  volume = {53},
  pages = {249--315},
  year = {1974},
  doi = {10.1002/sapm1974534249}
}

@article{Takhtajan1977,
  author = {L. A. Takhtajan},
  title = {Integration of the continuous Heisenberg spin chain through the inverse scattering method},
  journal = {Phys. Lett. A},
  volume = {64},
  pages = {235--237},
  year = {1977},
  doi = {10.1016/0375-9601(77)90727-7}
}

@article{Lakshmanan1977,
  author = {M. Lakshmanan},
  title = {Continuum spin system as an exactly solvable dynamical system},
  journal = {Phys. Lett. A},
  volume = {61},
  pages = {53--54},
  year = {1977},
  doi = {10.1016/0375-9601(77)90262-6}
}

@article{LakshmananPorsezian1990,
  author = {M. Lakshmanan and K. Porsezian},
  title = {Planar radially symmetric Heisenberg spin system and generalized nonlinear Schr\"odinger equation: Gauge equivalence, B\"acklund transformations and explicit solutions},
  journal = {Phys. Lett. A},
  volume = {146},
  pages = {329--334},
  year = {1990},
  doi = {10.1016/0375-9601(90)90964-P}
}

@book{AblowitzClarkson1991,
  author = {Mark J. Ablowitz and Peter A. Clarkson},
  title = {Solitons, Nonlinear Evolution Equations and Inverse Scattering},
  publisher = {Cambridge University Press},
  address = {Cambridge},
  year = {1991},
  doi = {10.1017/CBO9780511623998}
}

@book{Fokas2008,
  author = {Athanassios S. Fokas},
  title = {A Unified Approach to Boundary Value Problems},
  publisher = {SIAM},
  address = {Philadelphia},
  year = {2008},
  doi = {10.1137/1.9780898717068}
}

@article{BealsCoifman1984,
  author = {Richard Beals and Ronald R. Coifman},
  title = {Scattering and inverse scattering for first order systems},
  journal = {Commun. Pure Appl. Math.},
  volume = {37},
  pages = {39--90},
  year = {1984},
  doi = {10.1002/cpa.3160370105}
}

@article{MartinSiggiaRose1973,
  author = {P. C. Martin and E. D. Siggia and H. A. Rose},
  title = {Statistical dynamics of classical systems},
  journal = {Phys. Rev. A},
  volume = {8},
  pages = {423--437},
  year = {1973},
  doi = {10.1103/PhysRevA.8.423}
}

@article{OnsagerMachlup1953I,
  author = {Lars Onsager and Stefan Machlup},
  title = {Fluctuations and irreversible processes},
  journal = {Phys. Rev.},
  volume = {91},
  pages = {1505--1512},
  year = {1953},
  doi = {10.1103/PhysRev.91.1505}
}

@article{OnsagerMachlup1953II,
  author = {Stefan Machlup and Lars Onsager},
  title = {Fluctuations and irreversible processes. II. Systems with kinetic energy},
  journal = {Phys. Rev.},
  volume = {91},
  pages = {1512--1515},
  year = {1953},
  doi = {10.1103/PhysRev.91.1512}
}

@book{Kamenev2011,
  author = {Alex Kamenev},
  title = {Field Theory of Non-Equilibrium Systems},
  publisher = {Cambridge University Press},
  address = {Cambridge},
  year = {2011},
  doi = {10.1017/CBO9781139003667}
}

@article{SchorleppSasorovMeerson2023,
  author = {Timo Schorlepp and Pavel Sasorov and Baruch Meerson},
  title = {Short-time large deviations of the spatially averaged height of a Kardar--Parisi--Zhang interface on a ring},
  journal = {J. Stat. Mech.},
  volume = {2023},
  pages = {123202},
  year = {2023},
  doi = {10.1088/1742-5468/ad0a94}
}

@article{MeersonSasorov2014,
  author = {Baruch Meerson and Pavel V. Sasorov},
  title = {Extreme current fluctuations in lattice gases: Beyond nonequilibrium steady states},
  journal = {Phys. Rev. E},
  volume = {89},
  pages = {010101(R)},
  year = {2014},
  doi = {10.1103/PhysRevE.89.010101}
}

@article{PorsezianLakshmanan1991,
  author = {K. Porsezian and M. Lakshmanan},
  title = {On the dynamics of the radially symmetric Heisenberg ferromagnetic spin system},
  journal = {J. Math. Phys.},
  volume = {32},
  pages = {2923--2928},
  year = {1991},
  doi = {10.1063/1.529086}
}

@misc{Cantini2026,
  author = {Luigi Cantini},
  title = {An integrable approach to macroscopic fluctuation theory for the multispecies {SSEP}},
  year = {2026},
  eprint = {2606.27186},
  archivePrefix = {arXiv},
  primaryClass = {cond-mat.stat-mech}
}

@article{ScandiManzano2026,
  author = {Matteo Scandi and Gonzalo Manzano},
  title = {Universal Statistics of Charge Exchanges in Non-Abelian Quantum Transport},
  journal = {Phys. Rev. Lett.},
  volume = {136},
  pages = {150403},
  year = {2026},
  doi = {10.1103/n2gp-8bx9},
  eprint = {2508.15540},
  archivePrefix = {arXiv},
  primaryClass = {quant-ph}
}

@misc{SM,
  year = {2026},
  note = {See Supplemental Material.}
}

@article{BorceaBrandenLiggett2009,
  author = {Julius Borcea and Petter Br{\"a}nd{\'e}n and Thomas M. Liggett},
  title = {Negative dependence and the geometry of polynomials},
  journal = {J. Am. Math. Soc.},
  volume = {22},
  pages = {521--567},
  year = {2009},
  doi = {10.1090/S0894-0347-08-00618-8}
}

@article{ProlhacMallick2008,
  author = {Sylvain Prolhac and Kirone Mallick},
  title = {Current fluctuations in the exclusion process and Bethe Ansatz},
  journal = {J. Phys. A: Math. Theor.},
  volume = {41},
  pages = {175002},
  year = {2008},
  doi = {10.1088/1751-8113/41/17/175002}
}

@article{BodineauDerrida2025Perturbative,
  author = {Thierry Bodineau and Bernard Derrida},
  title = {A Perturbative Approach to the Macroscopic Fluctuation Theory},
  journal = {J. Stat. Phys.},
  volume = {192},
  pages = {51},
  year = {2025},
  doi = {10.1007/s10955-025-03439-4}
}

@misc{Derrida2025Lectures,
  author = {Bernard Derrida},
  title = {Lecture notes on large deviations in non-equilibrium diffusive systems},
  year = {2025},
  eprint = {2505.15618},
  archivePrefix = {arXiv},
  primaryClass = {math-ph}
}

@misc{Hurtado2025Lectures,
  author = {Pablo I. Hurtado},
  title = {Optimal paths and dynamical symmetry breaking in the current fluctuations of driven diffusive media},
  year = {2025},
  eprint = {2501.09629},
  archivePrefix = {arXiv},
  primaryClass = {cond-mat.stat-mech}
}

@misc{BertiniGabrielliJonaLasinio2025Schrodinger,
  author = {Lorenzo Bertini and Davide Gabrielli and Giovanni Jona-Lasinio},
  title = {Macroscopic fluctuation theory from a Lagrangian viewpoint and the Schr{\"o}dinger problem},
  year = {2025},
  eprint = {2506.22085},
  archivePrefix = {arXiv},
  primaryClass = {math.PR}
}

@article{Bressloff2025Targets,
  author = {Paul C. Bressloff},
  title = {Macroscopic fluctuation theory and the absorption of Brownian particles by partially reactive targets},
  journal = {J. Phys. A: Math. Theor.},
  volume = {58},
  pages = {505001},
  year = {2025},
  doi = {10.1088/1751-8121/ae26ff}
}

@article{BerliozBenichouGrabsch2025Driven,
  author = {Th{\'e}otim Berlioz and Olivier B{\'e}nichou and Aur{\'e}lien Grabsch},
  title = {Tracer and Current Fluctuations in Driven Diffusive Systems},
  journal = {Phys. Rev. Lett.},
  volume = {134},
  pages = {247101},
  year = {2025},
  doi = {10.1103/4j5q-j4ht}
}

@article{BerliozBenichouGrabsch2026Fick,
  author = {Th{\'e}otim Berlioz and Olivier B{\'e}nichou and Aur{\'e}lien Grabsch},
  title = {A shortcut through the macroscopic fluctuation theory: a generalised Fick law},
  journal = {J. Stat. Mech.},
  volume = {2026},
  pages = {063205},
  year = {2026},
  doi = {10.1088/1742-5468/ae76fe}
}

@article{GabrielliHarris2025,
  author = {Davide Gabrielli and Rosemary J. Harris},
  title = {Current Fluctuations for the Boundary-Driven Zero-Range Process on Graphs: Microscopic Versus Macroscopic Approach and a Theory of Non-reversible Resistor-Like Networks},
  journal = {Ann. Henri Poincar{\'e}},
  year = {2025},
  doi = {10.1007/s00023-025-01582-y},
  eprint = {2409.01337},
  archivePrefix = {arXiv}
}

@article{DandekarKrapivskyMallick2024Dyson,
  author = {Rahul Dandekar and P. L. Krapivsky and Kirone Mallick},
  title = {Current fluctuations in the Dyson gas},
  journal = {Phys. Rev. E},
  volume = {110},
  pages = {064153},
  year = {2024},
  doi = {10.1103/PhysRevE.110.064153}
}

@misc{SuzukiSasamoto2026,
  author = {Daisuke Suzuki and Tomohiro Sasamoto},
  title = {Non-stationary current fluctuations in 1D boundary-driven diffusive systems via Macroscopic Fluctuation Theory},
  year = {2026},
  eprint = {2605.27275},
  archivePrefix = {arXiv},
  primaryClass = {cond-mat.stat-mech}
}

@misc{JangidEtAl2026,
  author = {Sandeep Jangid and Soumyabrata Saha and Kapil Sharma and Jitendra Kethepalli and Benjamin Guiselin and Jacopo De Nardis and Tridib Sadhu},
  title = {An exactly solvable macroscopic fluctuation theory of single-file diffusion},
  year = {2026},
  eprint = {2607.14073},
  archivePrefix = {arXiv},
  primaryClass = {cond-mat.stat-mech}
}

@article{GrabschVenturelliBenichou2026,
  author = {Aur{\'e}lien Grabsch and Davide Venturelli and Olivier B{\'e}nichou},
  title = {Macroscopic fluctuation theory of interacting Brownian particles},
  journal = {Phys. Rev. E},
  volume = {113},
  pages = {054128},
  year = {2026},
  doi = {10.1103/1rvq-r3j1}
}

@misc{SahaJangidEtAl2026BottomUp,
  author = {Soumyabrata Saha and Sandeep Jangid and Thibaut Arnoulx de Pirey and Juliane U. Klamser and Tridib Sadhu},
  title = {A bottom-up approach to fluctuating hydrodynamics: Coarse-graining of stochastic lattice gases and the Dean--Kawasaki equation},
  year = {2026},
  eprint = {2601.02319},
  archivePrefix = {arXiv},
  primaryClass = {cond-mat.stat-mech}
}

@misc{SahaSadhu2025,
  author = {Soumyabrata Saha and Tridib Sadhu},
  title = {Large deviations of density in the non-equilibrium steady state of boundary-driven diffusive systems},
  year = {2025},
  eprint = {2501.03164},
  archivePrefix = {arXiv},
  primaryClass = {cond-mat.stat-mech}
}

@article{YoshimuraKrajnik2025,
  author = {Takato Yoshimura and {\v Z}iga Krajnik},
  title = {Anomalous current fluctuations from Euler hydrodynamics},
  journal = {Phys. Rev. E},
  volume = {111},
  pages = {024141},
  year = {2025},
  doi = {10.1103/PhysRevE.111.024141}
}

@article{YoshimuraKrajnik2026,
  author = {Takato Yoshimura and {\v Z}iga Krajnik},
  title = {Hydrodynamic fluctuations of stochastic charged cellular automata},
  journal = {Phys. Rev. E},
  volume = {113},
  pages = {064120},
  year = {2026},
  doi = {10.1103/sjxq-hclt}
}

@misc{AlbertEtAl2026QSSEP,
  author = {Mathias Albert and Denis Bernard and Tony Jin and Stefano Scopa and Shiyi Wei},
  title = {Universal classical and quantum fluctuations in the large deviations of current of noisy quantum systems: The case of {QSSEP} and {QSSIP}},
  year = {2026},
  eprint = {2601.16883},
  archivePrefix = {arXiv},
  primaryClass = {cond-mat.stat-mech}
}
\end{document}